\documentclass{svproc}
\usepackage[utf8]{inputenc}

\usepackage{url}

\usepackage{natbib}
\bibpunct{(}{)}{;}{a}{}{,}%
\usepackage{graphicx}
\usepackage{subcaption}
\usepackage{amsmath,amssymb}
\usepackage{multirow}
\usepackage{longtable}

\usepackage[b5paper]{geometry}
\begin{document}
	\mainmatter              
	\title{Optical light curves of light-weight supermassive black holes produced by the Zwicky Transient Facility Forced Photometry Service}
	\titlerunning{Optical light curves of light-weight SMBHs from ZTF}  
	%
	\author{Mariia Demianenko\inst{1, 2, 3} \and  Kirill Grishin\inst{2} \and Victoria Toptun\inst{2, 5} \and Igor Chilingarian\inst{4, 2} \and Ivan Katkov\inst{6, 7, 2} \and Vladimir Goradzhanov\inst{2, 5} \and Ivan Kuzmin\inst{2, 5}}
	\authorrunning{Demianenko et al.} 
	%
	\tocauthor{Mariia Demianenko, Igor Chilingarian, Kirill Grishin, Vladimir Goradzhanov, Victoria Toptun, Ivan Katkov, and Ivan Kuzmin}
	\institute{Moscow Institute of Physics and Technology (National Research University)\\
		\email{demyanenko.mv@phystech.edu},\\
		\and
		Sternberg Astronomical Institute, M.V. Lomonosov Moscow State University \\
		\and 
		HSE University \\
		\and
		Center for Astrophysics -- Harvard and Smithsonian (USA) \\
        \and
		Department of Physics, M.V. Lomonosov Moscow State University
		\and 
		New York University Abu Dhabi (UAE) \\
		\and 
		Center for Astro, Particle, and Planetary Physics, NYU Abu Dhabi (UAE)}
	\maketitle 
\begin{abstract}
	In this paper, we present an algorithm to correct optical light curves obtained using The Zwicky Transient Facility Forced Photometry Service and its application to the analysis of optical variability of 136 actvie galactic nuclei (AGN) powered by ``light-weight'' supermassive black holes (SMBH; $M_{BH}<2*10^6 \sun$) including 24 intermediate-mass black holes (IMBH; $M_{BH}<2*10^5 \sun$). We detected variability in nearly all sources and also analyzed its dependence on the X-ray luminosity for 101 objects. We also identified a previously unknown candidate tidal disruption event (TDE) in SDSS~J112637.74+513423.0.
\end{abstract}

\paragraph{\textbf{Introduction.}}
The presence of optical variability in a galaxy center is one of the most reliable diagnostics of an AGN to date.
Optical light curves are used to search for a statistically significant variability of emission. 
The size of a region illuminated by an AGN in optical wavelengths defines the stochastic variability time scale in the light curves of such objects according to e.g. \citet{2021Sci...373..789B}. This allows one to use optical light curves to search and confirm IMBHs (\citealp{2020ApJ...889..113M}) because $M_{BH}$ limits the AGN luminosity. $M_{BH}$ can be measured from broad optical emission line profiles in AGN spectra \citep{2004ApJ...610..722G,optic_fm,2018ApJ...863....1C}.

\begin{figure}
\includegraphics[width=0.48\hsize]{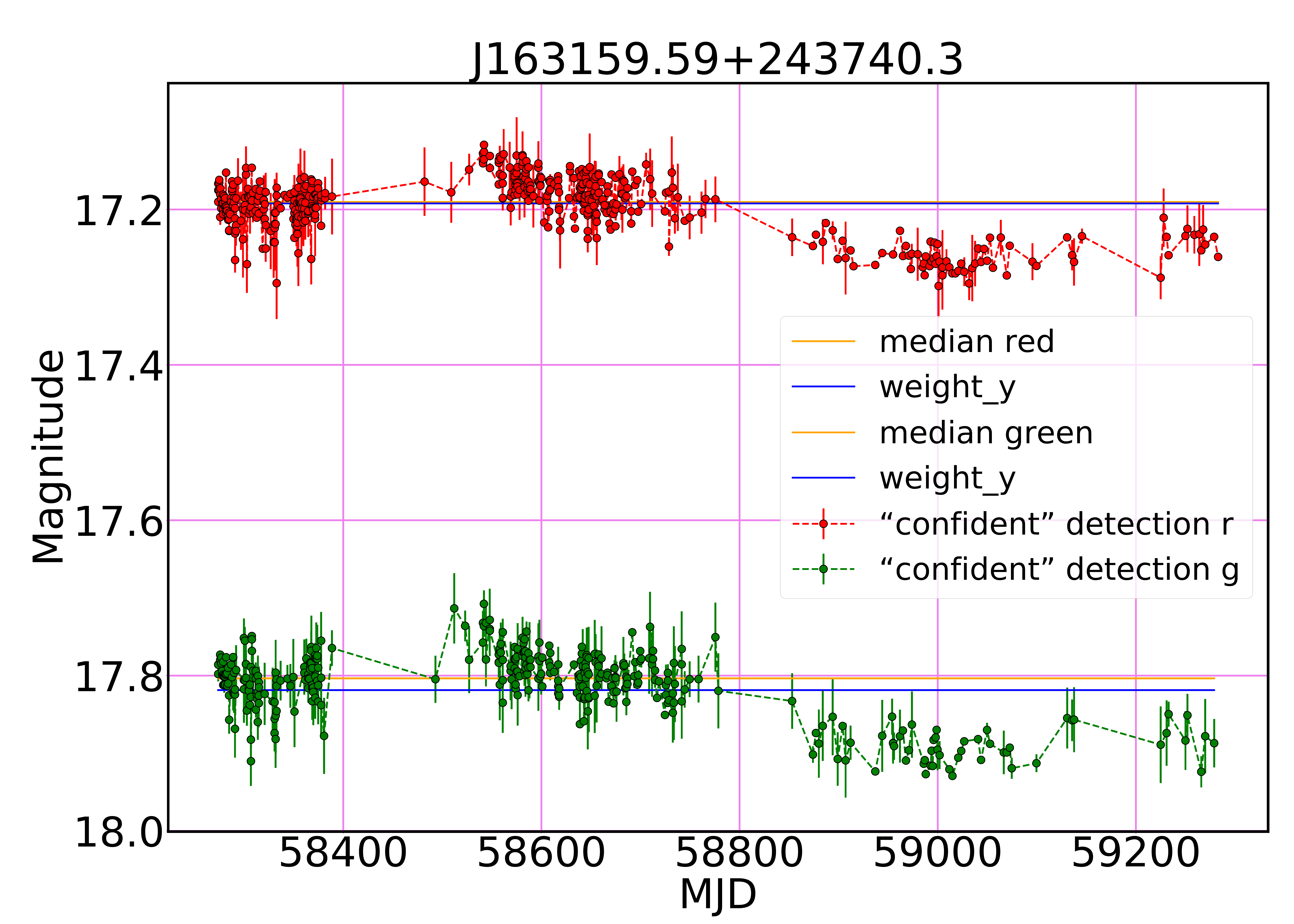}
\includegraphics[width=0.48\hsize]{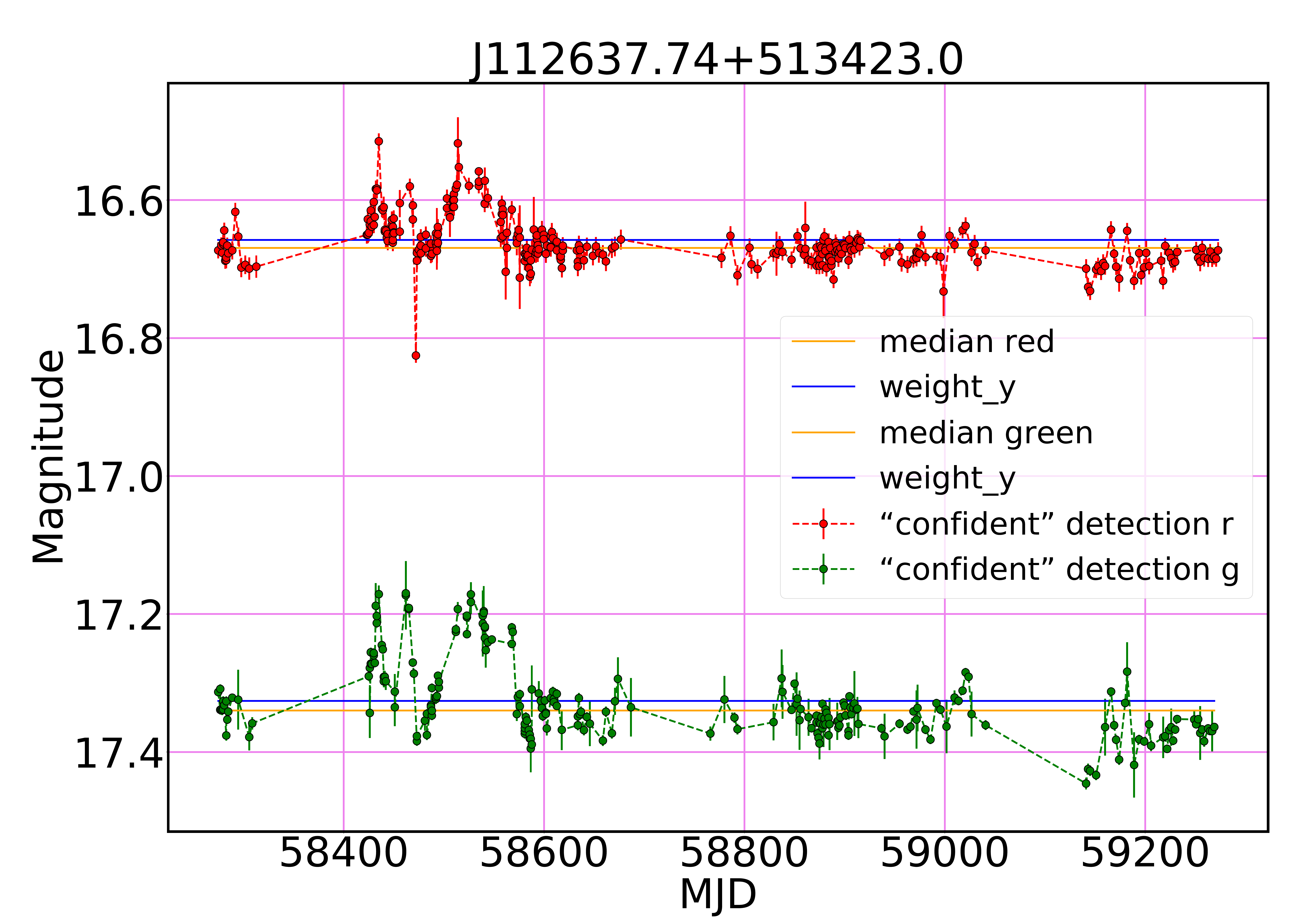}\\
\vskip -5mm
\includegraphics[width=0.48\hsize]{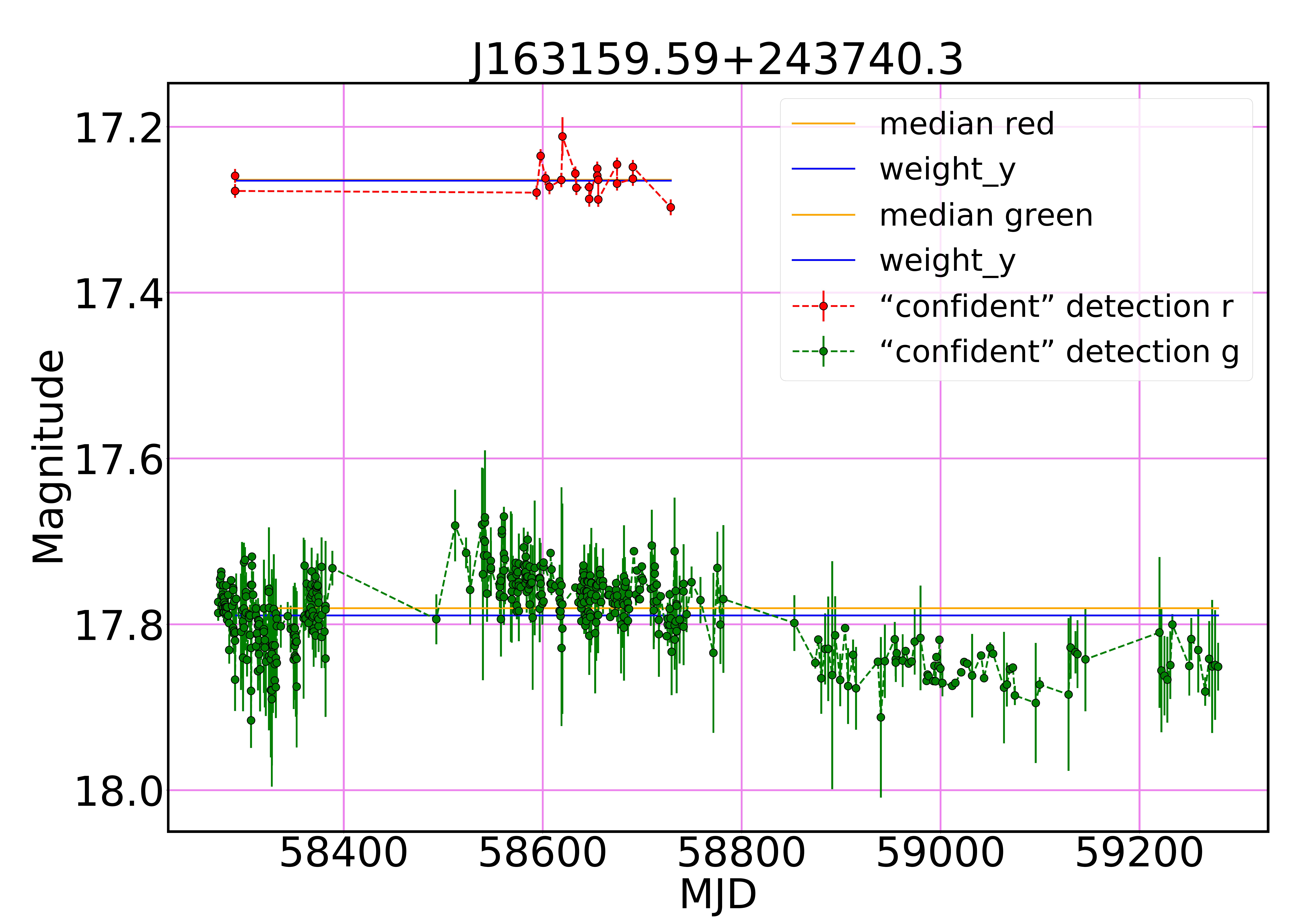}
\includegraphics[trim=-0.0cm 0 0.0cm 0,width=0.48\hsize]{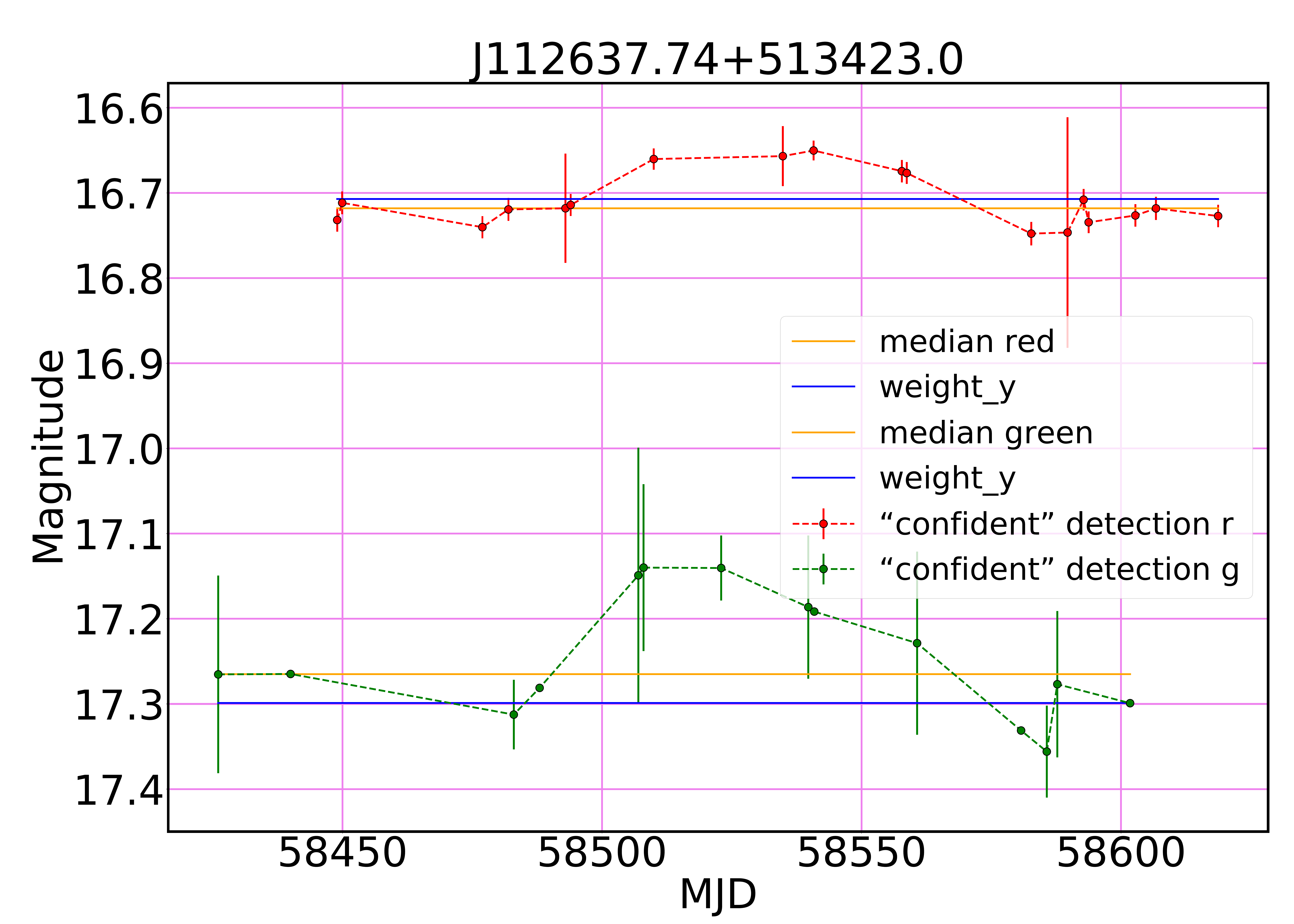}
\caption{\small{{\bf Top left:} The light curve J163159.59+243740.3 obtained using ZTF Forced Photometry Service. {\bf Top right:} The light curve of the candidate TDE before correction. {\bf Bottom left:} The light curve of J163159.59+243740.3 after correction, points in one field of one quadrant of one CCD are visualized. {\bf Bottom right:} The light curve of the candidate TDE after correction. The points in the most occurring quadrant of the CCD are shown.\label{fig_lc}}}
\end{figure}

\begin{figure}
    \centering
\vskip -5mm
      {\includegraphics[width=0.48\linewidth]{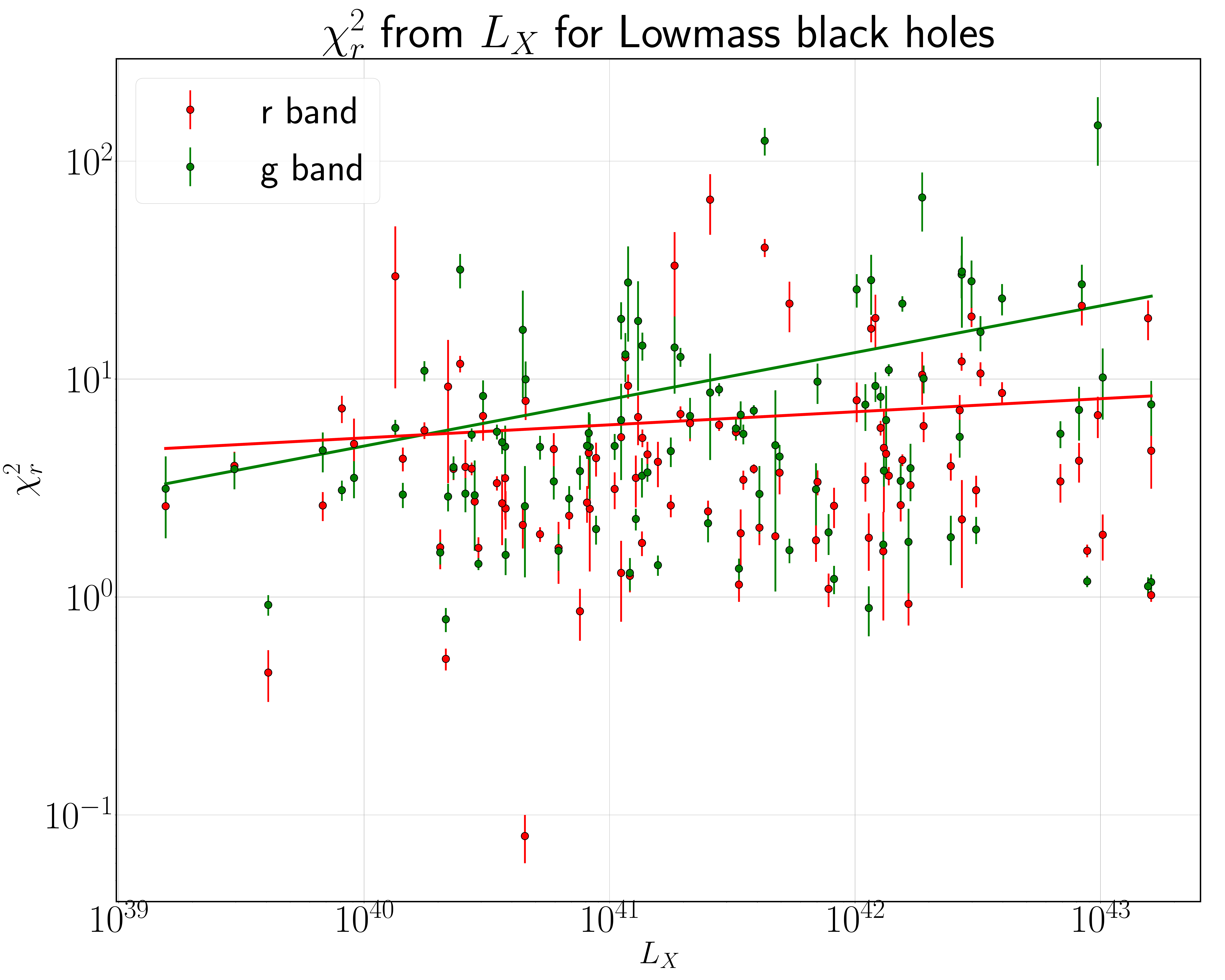}}
    \caption{\small{Dependence of the the optical light curve variability amplitude on the X-ray luminosity the two ZTF photometric bands, `ZTF g' (green) and `ZTF r' (red).}}
    \label{fig_var}
\end{figure}

\paragraph{\textbf{Data.}}
\citet{2018ApJ...863....1C} defined a sample of IMBHs from the analysis of optical spectra with the {\sc NBursts} full spectral fitting \citep{2007IAUS..241..175C,2007MNRAS.376.1033C} in the RCSED catalog \citep{2017ApJS..228...14C}, with $M_{BH}$ estimated between $3*10^4 M\sun$ and $2*10^5 M\sun$. 10 of these objects were confirmed in X-ray in \citet{2018ApJ...863....1C}, and another 14 were confirmed later from new and archival X-ray data. We define light-weight SMBHs as SMBHs having $M_{BH}<2*10^6 M \sun$.
112 objects from the expanded sample of candidates light-weight SMBHs were confirmed by archival X-ray data. In total, we studies 136 optical light curves for 24 \textit{bona fide} IMBHs and 112 \textit{bona fide} light-weight SMBHs. Out of 136 objects, 101 had photometric points in two filters, `ZTF r' and 'ZTF g'. 

\paragraph{\textbf{Analysis of optical variability in light-weight SMBHs and IMBHs.}}
To estimate the optical variability amplitude and its timescale, we need to construct calibrated light curves without systematic variations exceeding 10-15\% of the host galaxy flux. We use differential images, which are obtained by subtracting the reference image from a current science image matching the point-spread-function (PSF). The reference image is a weighted average of all images for a given field, degraded to the same resolution using a Gaussian filter. Then, PSF photometry is performed without restrictions on the flux positivity. This approach is implemented in the ZTF Forced Photometry Service.

The resulting points do not always have the same zero point, because they were observed in different quadrants of different CCDs in the focal plane of the telescope. And also, color corrections are performed using zero color from the PanSTARRS photometric system. This correction is well defined only for those objects for which zero color measurement is available, otherwise the correction is estimated by the interpolation of PanSTARRS values. For each quadrant of the CCD and the field, the higher and lower flux limits of that sensor are defined, and points with flux measurements outside these limit derived from a specific observation should be excluded \citep{2019PASP..131a8003M}.
Here, we set the threshold signal-to-noise ratio (SNR) to 3. 
The distance to the nearest reference was set to $<$1~arcsec, pixel quality indicators helped to clear outliers related to uncorrected cosmic ray hits and hot pixels. The zero-point correction was made for each observation, based on the CCD quadrant, and the average airmass during an observation. Then we defined the level of significance according to the $\chi^2$ criterion at which the light curve is not consistent with white noise on top of a constant. Many of the studied objects showed statistically significant variability on timescales from days to months.

Two examples of uncorrected ZTF Forced Photometry light curves are shown in Fig.~\ref{fig_lc} (top row). 
The bottom row of Fig.~\ref{fig_lc} shows the light curves after correction. 
In Fig.\ref{fig_var} we show the dependence of the amplitude of optical variability (errors calculated using bootstrap) from the X-ray luminosity for 101 objects from our sample. 

Also, a strong flare with an amplitude of 0.25 mag was observed in the object J112637.74+513423.0 for about 3 months, which can be either a heavily dust obscured supernova explosion or a star ruptured by tidal forces in the SMBH vicinity (TDE). 
We show a ZTF light curve of the TDE candidate in Fig~\ref{fig_lc} (top right).
A more reliable light curve after the correction is shown in Fig~\ref{fig_lc} (bottom right): the ``hump'' is reliably traced by data points even in one CCD quadrant hence clearly confirming that the flare is real.

The bottleneck that limits the scalability of our analysis and makes it yet impossible to study thousands of AGN is that the ZTF Forced Photometry Service can process up to 100 objects at a time and it takes hours to weeks to process every request.  

\paragraph{\textbf{Conclusions.}}
The use of optical variability confirmed the high quality of the selection of IMBH and light-weight SMBHs candidates using a broad component $H\alpha$ and, in the future, will allow us to exclude candidates where broad lines have a transient or star formation related nature. Also, light curve analysis can be used for independent confirmation or selection of IMBH and light-weight SMBH candidates, because there are strong correlations of $M_{BH}$ and the optical timescale according to \citet{2021Sci...373..789B}.
\textbf{Acknowledgement.}
MD is supported by the RSF project 17-72-20119.
\vskip 5mm
\begingroup
\let\clearpage\relax
\bibliographystyle{aa}
\bibliography{ZTF_IMBH}
\endgroup

\end{document}